\documentclass[
    ,final            
  ]
  {aipproc}

\layoutstyle{8x11single}

\newcommand{\pvec}{\mathbf{p}}
\newcommand{\xvec}{\mathbf{x}}
\newcommand{\yvec}{\mathbf{y}}

\begin{document}

\title{The excited hadron spectrum in lattice QCD using a new method of 
       estimating quark propagation}


\author{C.~Morningstar}{
  address={Dept.~of Physics, Carnegie Mellon University, 
           Pittsburgh, PA 15213, USA}
}

\author{A.~Bell}{
  address={Dept.~of Physics, Carnegie Mellon University, 
           Pittsburgh, PA 15213, USA}
}

\author{J.~Bulava}{
  address={NIC, DESY, Platanenallee 6, D-15738, Zeuthen, Germany}
}

\author{E.~Engelson}{
  address={Dept.~of Physics, University of Maryland, 
           College Park, MD 20742, USA}
}

\author{J.~Foley}{
  address={Dept.~of Physics, Carnegie Mellon University, 
           Pittsburgh, PA 15213, USA}
}

\author{K.J.~Juge}{
  address={Dept.~of Physics, University of the Pacific, Stockton, CA 95211, USA}
}

\author{D.~Lenkner}{
  address={Dept.~of Physics, Carnegie Mellon University, 
           Pittsburgh, PA 15213, USA}
}

\author{M.~Peardon}{
  address={School of Mathematics, Trinity College, Dublin 2, Ireland}
}

\author{S.~Wallace}{
  address={Dept.~of Physics, University of Maryland, 
           College Park, MD 20742, USA}
}

\author{C.H.~Wong}{
  address={Dept.~of Physics, Carnegie Mellon University, 
           Pittsburgh, PA 15213, USA}
}

\begin{abstract}
 Progress in determining the spectrum of excited baryons and mesons in lattice
 QCD is described.  Large sets of carefully-designed hadron operators have been
 studied and their effectiveness in facilitating the extraction of excited-state
 energies is demonstrated.  A new method of stochastically estimating the
 low-lying effects of quark propagation is proposed which will allow reliable
 determinations of temporal correlations of single-hadron and multi-hadron 
 operators.
\end{abstract}

\maketitle


To extract excited-state energies in Monte Carlo calculations of lattice QCD,
correlation matrices must be evaluated and operators which couple well to the
states of interest are crucial.  To study a particular state of interest, all 
states lying below that state must first be extracted, and as the pion gets lighter 
in lattice QCD simulations, more and more multi-hadron states will lie below the 
excited resonances.  To reliably determine the energies of the multi-hadron states,
multi-hadron operators made from constituent hadron operators with well-defined
relative momenta will most likely be needed, and the computation of temporal 
correlation functions involving such operators requires accurately including
the low-lying effects of propagation from all spatial sites on one time slice
to all sites on another (or the same) time slice.  This talk is a progress
report on our efforts to study the excited-state spectrum of QCD using the
Monte Carlo method: results from our process of selecting optimal single-hadron
operators are presented, and a new method of estimating slice-to-slice quark
propagators is proposed.

The use of operators whose correlation functions $C(t)$ attain their
asymptotic form as quickly as possible is crucial for reliably
extracting excited hadron masses.  An important ingredient in constructing
such hadron operators is the use of smeared fields.  Operators constructed
from smeared fields have dramatically reduced mixings with the high frequency
modes of the theory.  Both link-smearing and quark-field smearing must
be applied.  Since excited hadrons are expected to be large objects, 
the use of spatially extended operators is another key ingredient in
the operator design and implementation.  A more detailed discussion
of these issues can be found in Ref.~\cite{baryons2005A}.


A first glimpse of the nucleon excitation spectrum using two flavors of 
dynamical quarks was presented in Ref.~\cite{spectrum2009}.
Highly excited states for isospin 1/2 baryons 
were calculated for the first time using lattice QCD with unquenched
configurations.  Results were obtained for two pion masses: 416(36)~MeV
and 578(29)~MeV. The lowest four energies were reported in each of the 
six irreducible representations of the octahedral group at each pion 
mass.  Clear evidence was found for a $5/2^-$ state in the pattern of 
negative-parity excited states. This agrees with the pattern of 
physical states and spin 5/2 has been realized for the first time in
lattice QCD.  Note that in this study, only single-particle
interpolating operators were used, so the use of \textit{point-to-all} 
quark propagators was adequate.

\begin{figure}
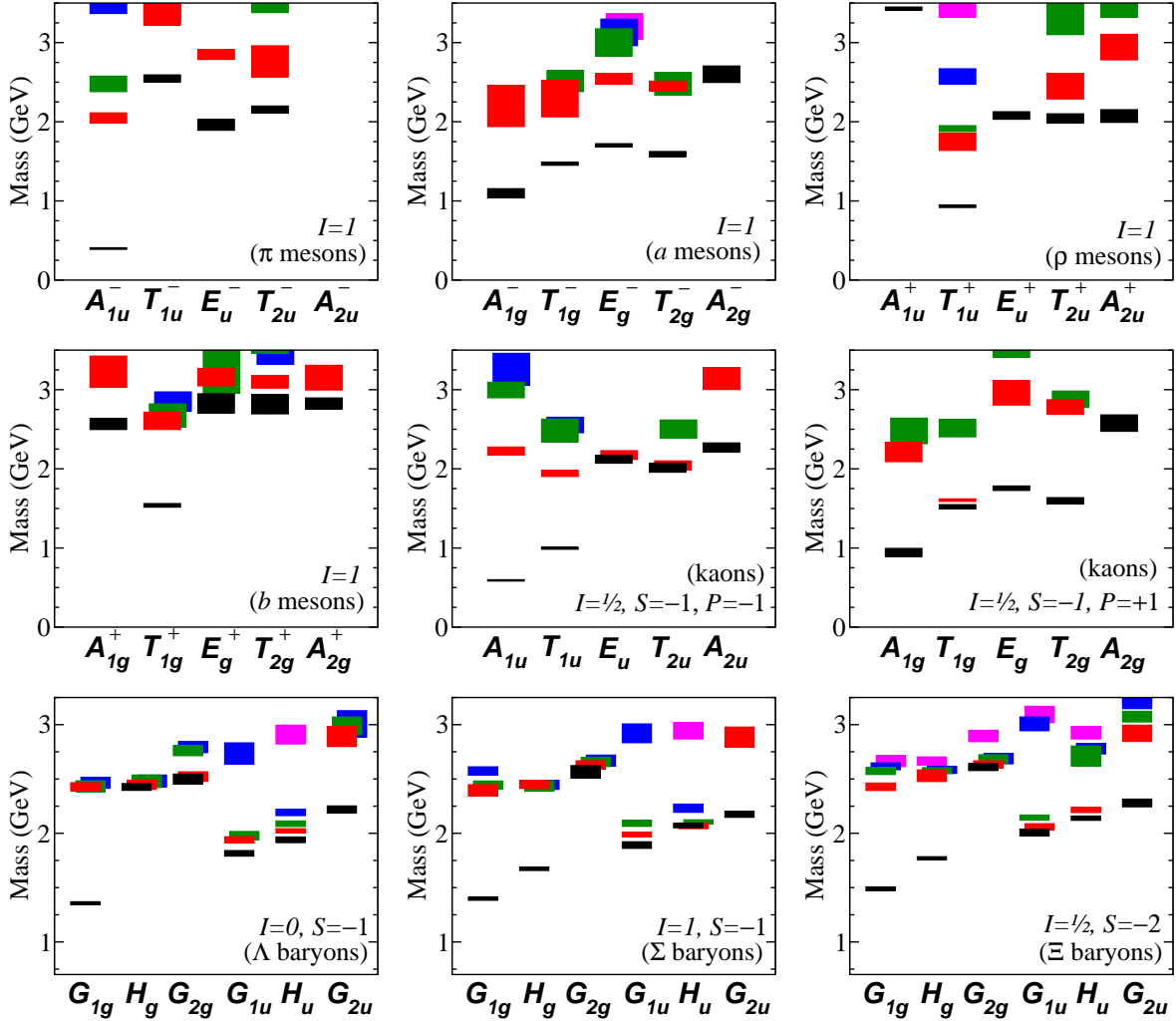

\begin{minipage}{6.0in}
\hspace*{-4mm}
\includegraphics[width=1.95in,bb=31 34 523 461]{pi_mesons_pruning.eps}\quad
\includegraphics[width=1.95in,bb=31 34 523 461]{a_mesons_pruning.eps}\quad
\includegraphics[width=1.95in,bb=31 34 523 461]{rho_mesons_pruning.eps}\\[3mm]
\hspace*{-4mm}
\includegraphics[width=1.95in,bb=31 34 523 461]{b_mesons_pruning.eps}\quad
\includegraphics[width=1.95in,bb=31 34 523 461]{kaons_oddp_pruning.eps}\quad
\includegraphics[width=1.95in,bb=31 34 523 461]{kaons_evenp_pruning.eps}\\[3mm]
\hspace*{-4mm}
\includegraphics[width=1.95in,bb=31 34 523 461]{lambda_baryons_pruning.eps}\quad
\includegraphics[width=1.95in,bb=31 34 523 461]{sigma_baryons_pruning.eps}\quad
\includegraphics[width=1.95in,bb=31 34 523 461]{xi_baryons_pruning.eps}
\end{minipage}
\caption{
Hadron operator selection: low-statistics simulations have been performed
to study the hundreds of single-hadron operators produced by our group-theoretical
construction.  A ``pruning" procedure was followed in each channel to select
good sets of between six to a dozen operators.  The plots above show the
stationary-state energies extracted to date from correlation matrices of the finally
selected single-hadron operators.  Results were obtained using between 50 to 100 
configurations on a $16^3\times 128$ anisotropic lattice for $N_f=2+1$ quark flavors
with spacing $a_s\sim 0.12$~fm, $a_s/a_t\sim 3.5$, and quark masses such that
$m_\pi\sim 380$~MeV.  Each box indicates the energy of one stationary 
state; the vertical height of each box indicates the statistical error.
\label{fig:pruning}}
\end{figure}

A large effort was undertaken during the summer of 2009 to select optimal
sets of baryon and meson operators in a large variety of isospin sectors.
The following procedure was used. (1) First, operators with excessive 
intrinsic noise were removed.  This was done by examining the diagonal 
elements of the correlation matrix and discarding
those operators whose self-correlators had relative errors above some
threshold for a range of temporal separations.  
(2) Second, pruning within operator types (single-site, singly-displaced, 
\textit{etc.}) was done based on the condition number of the submatrices
$\widehat{C}(t=a_t)$, where
$\widehat{C}_{ij}(t) = C_{ij}(t)/\sqrt{C_{ii}(t)C_{jj}(t)}$ with
$C_{ij}(t)=\langle 0\vert\ O_i(t)O^\dagger_j(0)\ \vert 0\rangle$.
The condition number was taken to be the ratio of the largest eigenvalue
over the smallest eigenvalue.  A value near unity is ideal, indicating
near orthogonality of the states produced by the action of the operators
on the vacuum.  Orthogonality prevents noise from creeping into the 
eigenvalues of the correlation matrix.
For each operator type, the set of about six operators which yielded
the lowest condition number of the above submatrix was retained.
(3) Lastly, pruning across all operator types was done based again on 
the condition number of the remaining submatrix as defined above.  In this
last step, the goal was to choose between 8 to 12 operators, keeping two or three
of each type, such that a condition number reasonably close to unity was
obtained.  As long as a good variety of operators was retained, the 
resulting spectrum seemed to be fairly independent of the exact choice of
operators at this stage. 

Low-statistics Monte Carlo computations were done to accomplish the above
operator selections using between 50 to 100 configurations on a $16^3\times 128$ 
anisotropic lattice for $N_f=2+1$ quark flavors with spacing $a_s\sim 0.12$~fm, 
$a_s/a_t\sim 3.5$, and quark masses such that the pion has mass around 380~MeV. 
The method described in Ref.~\cite{distillation2009} was used.
Stationary-state energies using the finally selected operator sets are shown in
Fig.~\ref{fig:pruning}.  The nucleon, $\Delta$, $\Xi$, $\Sigma$, and $\Lambda$
baryons were studied, and light isovector and kaon mesons were investigated. 
Hundreds of operators were studied, and optimal sets containing eight or so
operators in each symmetry channel were found.  Future computations will focus
solely on the operators in the optimal sets. 

\begin{figure}
\includegraphics[width=3.5in,bb=0 30 743 324]{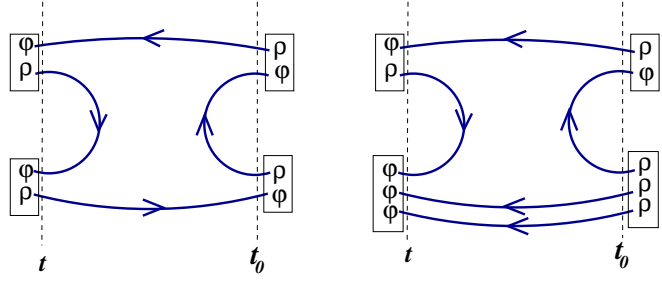}
\caption{
Diagrams of multi-hadron correlators that require having $\varrho$ 
noise sources on the later time $t$. Solution vectors are denoted
by $\varphi$. (Left) A two-meson correlator. (Right) The correlator
of a baryon-meson system.
\label{fig:multicorr}}
\end{figure}

To study a particular eigenstate of interest in the Monte Carlo method, all 
eigenstates lying below that state must first be extracted, and as the pion gets
lighter in lattice QCD simulations, more and more \textit{multi-hadron} states will
lie below the excited resonances.  A \textit{good} baryon-meson operator of total
zero momentum is typically a superposition of terms having the form
\[
 B(\pvec,t)M(-\pvec,t)=\frac{1}{V^2}\sum_{\xvec,\yvec}\varphi_B(\xvec,t)
\varphi_M(\yvec,t)e^{i\pvec\cdot(\xvec-\yvec)},
\]
where $V$ is the spatial volume of the lattice, $\pvec$ is the relative
momentum, and $\varphi_B(\xvec,t)$ and $\varphi_M(\yvec,t)$ are appropriate localized 
interpolating fields for a baryon and a meson, respectively.  In the evaluation of 
the temporal correlations of such a multi-hadron operator, it is not possible to
completely remove all summations over the spatial sites on the source time-slice using
translation invariance.  Hence, the need for estimates of the quark propagators from 
all spatial sites on a time slice to all spatial sites on another time slice
cannot be sidestepped.  Some correlators will involve diagrams with
quark lines originating at the sink time $t$ and terminating at the
same sink time $t$ (see Fig.~\ref{fig:multicorr}), so quark propagators involving
a large number of starting times $t$ must also be handled.

Finding better ways to stochastically estimate such slice-to-slice quark 
propagators for large numbers of quark-line starting times
is crucial to the success of our excited-state hadron spectrum project
at lighter pion masses.  During the past two years, we have been exploring and 
developing new methods for dealing with such propagators. A first new method, 
known as distillation\cite{distillation2009}, was devised in the summer of 2008. 
We found that direct calculations with this method are feasible on small lattices,
but the cost of the computations using this method rises rapidly with the
spatial volume.  During the past year, the quark-field smearing used in the
distillation method was then combined with a stochastic approach to estimating 
the quark propagators, resulting in a method suitable for large volumes.

The new quark-field smearing scheme, here called Laplacian Heaviside (Laph),
has been described in Ref.~\cite{distillation2009} and is defined by
\begin{equation}
\widetilde{\psi}(x) = 
 \Theta\left(\sigma_s^2+\widetilde{\Delta}\right)\psi(x),
\end{equation}
where $\widetilde{\Delta}$ is the three-dimensional covariant Laplacian
in terms of the stout-smeared gauge field and $\sigma_s$ is the smearing cutoff
parameter.  The Heaviside function truncates the sum over Laplacian
eigenmodes, restricting the summation to some number $N_v$ of the lowest-lying
$-\widetilde{\Delta}$ eigenmodes.  The number of eigenvectors required increases
linearly with the spatial volume of the lattice.
In Ref.~\cite{distillation2009}, it was
demonstrated that this new smearing method is equally effective at reducing
excited-state contamination compared to prior smearing schemes, but this new
smearing is advantageous in that it simplifies the computation
of slice-to-slice quark propagators.  

Random noise vectors $\eta$ whose expectations satisfy
$E(\eta_i)=0$ and $E(\eta_i\eta_j^\ast)=\delta_{ij}$ are useful for 
stochastically estimating the inverse of a large matrix $M$ 
as follows.  Assume that for each of $N_R$ 
noise vectors, we can solve the following
linear system of equations: $M X^{(r)}=\eta^{(r)}$ for $X^{(r)}$.
Then $X^{(r)}=M^{-1}\eta^{(r)}$, and $E( X_i \eta_j^\ast ) = M^{-1}_{ij}$
so that a Monte Carlo estimate of $M_{ij}^{-1}$ is given by
$
  M_{ij}^{-1} \approx \lim_{N_R\rightarrow\infty} N_R^{-1}
 \sum_{r=1}^{N_R} X_i^{(r)}\eta_j^{(r)\ast}.
$
Unfortunately, this equation usually produces stochastic estimates with 
variances which are much too large to be useful.  Variance reduction is
done by \textit{diluting} the noise vectors\cite{alltoall}.
A given dilution scheme can be viewed as the application of a complete
set of projection operators $P^{(a)}$. Define
$
  \eta^{[a]}_k=P^{(a)}_{kk^\prime}\eta_{k^\prime} ,
$
and further define $X^{[a]}$ as the solution of
$
   M_{ik}X^{[a]}_k=\eta^{[a]}_i,
$
then we have\\[-3mm]
\begin{equation}
   M_{ij}^{-1}\approx \lim_{N_R\rightarrow\infty}\frac{1}{N_R}
 \sum_{r=1}^{N_R} \sum_a X^{(r)[a]}_i\eta^{(r)[a]\ast}_j.
\label{eq:diluted}
\end{equation}\\[-3mm]
The dilution projections ensure \textit{exact zeros} (zero variance) for our 
Monte Carlo estimates of many of the off-diagonal elements instead of 
estimates that are only statistically zero, resulting in a dramatic reduction
in the variance of the $M^{-1}$ estimates.  The use of $Z_4$ noise ensures
zero variance in our estimates of the diagonal elements.  The effectiveness
of the variance reduction depends on the projectors chosen.  

All previous stochastic methods have introduced noise in the full 
spin-color-space-time vector space, that is, on the entire lattice itself.  
However, since we intend to use Laplacian 
Heaviside quark-field smearing, an alternative is possible: noise vectors 
$\varrho$ can be introduced \textit{only in the Laph subspace}.  The noise
vectors $\varrho$ now have spin, time, and Laplacian eigenmode number
as their indices.  Color and space indices get replaced by Laplacian
eigenmode number.  Again, each component of $\varrho$ is a random
$Z_4$ variable so that $E(\varrho)=0$ and $E(\varrho\varrho^\dagger)=I_d$.
Dilution projectors $P^{(a)}$ can once again be introduced, but these
projectors are matrices in the Laph subspace.  Our dilution projectors are
products of time dilution, spin dilution, and Laph eigenvector dilution
projectors.  For each type (time, spin, Laph eigenvector) of dilution, we
studied four different dilution schemes.  Let $N$ denote the dimension
of the space of the dilution type of interest.  For time dilution, $N=N_t$
is the number of time slices on the lattice.  For spin dilution, $N=4$ is
the number of Dirac spin components.  For Laph eigenvector dilution, $N=N_v$
is the number of eigenvectors retained.  The four schemes we studied
are defined below:
\[ \begin{array}{lll}
P^{(a)}_{ij} = \delta_{ij},              & a=0,& \mbox{(no dilution)} \\
P^{(a)}_{ij} = \delta_{ij}\ \delta_{ai},  & a=0,\dots,N-1 & \mbox{(full dilution)}\\
P^{(a)}_{ij} = \delta_{ij}\ \delta_{a,\, \lfloor Ki/N\rfloor}& a=0,\dots,K-1, & \mbox{(block-$K$)}\\
P^{(a)}_{ij} = \delta_{ij}\ \delta_{a,\, i\bmod K} & a=0,\dots,K-1, & \mbox{(interlace-$K$)}
\end{array}\]
where $i,j=0,\dots,N-1$, and we assume $N/K$ is an integer.  We use a triplet
(T, S, L) to specify a given dilution scheme, where ``T" denote time,
``S" denotes spin, and ``L" denotes Laph eigenvector dilution.  The schemes
are denoted by 1 for no dilution, F for full dilution, and B$K$ and I$K$ for
block-$K$ and interlace-$K$, respectively.  For example, full time and spin
dilution with interlace-8 Laph eigenvector dilution is denoted by
(TF, SF, LI8).

\begin{figure}[t]
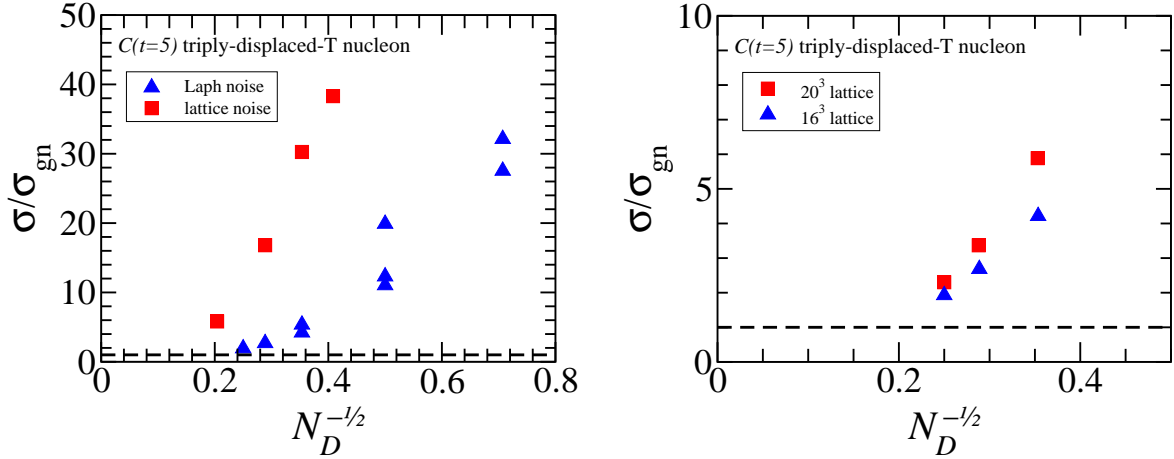

\includegraphics[width=3.0in,bb=17 32 573 454]{lattice_vs_laph_noise2.eps}
\hspace*{4mm}
\includegraphics[width=3.0in,bb=17 32 574 454]{volume_dependence2.eps}
\caption{(Left) Comparison of the new stochastic Laph method (triangles)
with the earlier stochastic method (squares) using noise on the
full lattice for the correlator $C(t=5a_t)$ of a
triply-displaced-T nucleon operator on a $16^3\times 128$ lattice.
The vertical scale is the ratio of statistical error $\sigma$ over the
error in the gauge-noise limit $\sigma_{\rm gn}$ (averaging over the six
permutations of the three noises is not done in the results shown in
these plots), and in the horizontal
scale, $N_D$ is the number of Dirac-matrix inversions per source per
quark line.  Each point shows an error ratio using a particular dilution
scheme. (Right) Comparison of the new Laph stochastic method on $16^3$ 
(triangles) and $20^3$ (squares) lattices.  The number of Laplacian 
eigenvectors needed is 32 on the $16^3$ lattice and 64 on the $20^3$ lattice.}
\label{fig:laph}
\end{figure}

Introducing noise in this
way produces correlation functions with significantly reduced variances,
as shown in Fig.~\ref{fig:laph}.  Let $C(t)$ denote the correlation function
of a representative triply-displaced-T nucleon operator at temporal separation 
$t$.  Let $\sigma_{\rm gn}$ represent the statistical error in $C(t=5a_t)$ using
exactly-determined slice-to-slice quark propagators.  In other words, 
$\sigma_{\rm gn}$ arises solely from the statistical fluctuations in the gauge
configurations themselves (the gauge noise).  Let $\sigma$ denote the error
in $C(t=5a_t)$ using stochastic estimates of the quark propagators.  The vertical 
axis in each plot of Fig.~\ref{fig:laph} is the ratio of the statistical error 
$\sigma$ in $C(t=5a_t)$ over $\sigma_{\rm gn}$.  Results are shown for a variety of 
different dilution schemes.  In the left plot, the squares show results for dilution
schemes with noise introduced in the larger spin-color-space-time vector
space, and the triangles show results for different dilution schemes
using noise introduced only in the Laph subspace.  One sees nearly an
order of magnitude reduction in the statistical error.

The volume dependence of this new method was found to be very mild.  Calculations
on a $16^3$ and a $20^3$ lattice were carried out and it was found that the ratio
$\sigma/\sigma_{\rm gn}$ of the correlator error over its error in the gauge-noise
limit only grew marginally (about 30\%) in the larger volume for the \textit{same}
number of inversions of the Dirac matrix.  Keep in mind that the number of Laplacian
eigenvectors needed doubles in going from the smaller to the larger volume.
The error ratios for the representative triply-displaced-T nucleon correlator
on a $16^3$ lattice (triangles) are compared to those from a $20^3$ lattice
(squares) in Fig.~\ref{fig:laph}.


\begin{figure}[t]
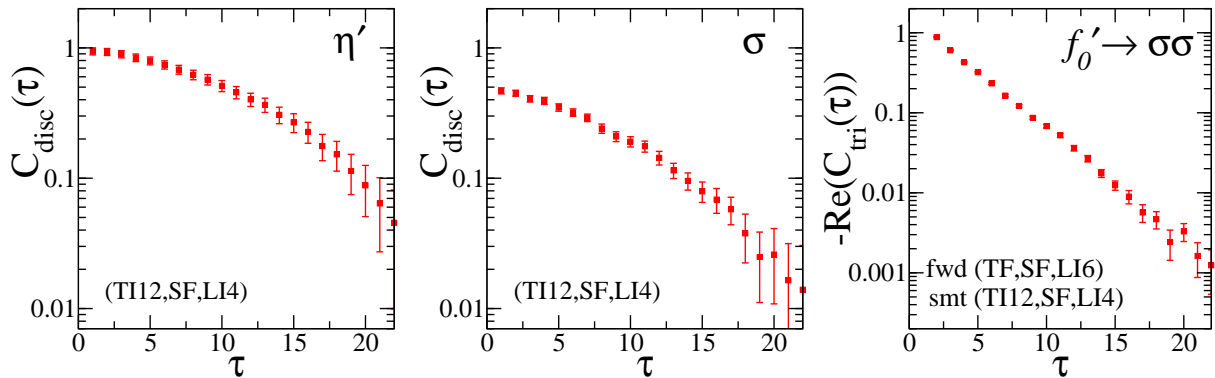

\includegraphics[width=2.0in,bb=15 54 526 523]{etaprime.eps}\quad
\includegraphics[width=2.0in,bb=15 54 526 523]{sigma.eps}\quad
\includegraphics[width=2.0in,bb=15 54 526 523]{f0sigsig.eps}
\caption{
Contributions to three correlators involving same-time quark lines evaluated 
using the new stochastic Laph method.  Results are obtained using 99 configs on a 
small $12^3\times 96$ anisotropic lattice for quark masses $m_u=m_d=m_s$.  
Quark lines from the source 
time $t_0$ to the sink time $t$ use full time and spin dilution and
interlace-6 Laph eigenvector dilution.  Quark lines from $t$ to $t$ use
interlace-12 time dilution, full spin dilution, and interlace-4 Laph
eigenvector dilution.  Contributions from only the disconnected diagram 
are shown for the $\eta^\prime$ (left) and $\sigma$ (center) correlators. (Right)
The contribution to the $f_0^\prime -\sigma\sigma$ off-diagonal correlator
from the two triangle diagrams only is shown. 
\label{fig:disc}}
\end{figure}

Different dilution schemes have been explored, and we have found that
the combination of full time and spin dilution (one projector for each component)
with interlace-8 Laph eigenvector dilution produces a variance near that of
the gauge noise limit.  Correlation matrix elements evaluated with this
method have sufficiently small variances so that diagonalization can be
reliably achieved.  Using interlace-12
time dilution with full spin and interlace-4 Laph eigenvector dilution, even
mixing diagrams between single-hadron and two-hadron states can be accurately
estimated, as shown in Fig.~\ref{fig:disc}.
The rightmost plot demonstrates that evaluating correlation functions
involving our multi-hadron operators will be feasible.

The next steps in our spectrum project are to carry out the operator selection
process for single mesons and baryons having non-zero momenta, combine them
to form multi-hadron operators, then complete 
computations of QCD stationary-state energies using, for the first time, both 
single-hadron and multi-hadron operators.
This work was supported by the U.S.~National Science Foundation 
under awards PHY-0510020, PHY-0653315, PHY-0704171 and through TeraGrid 
resources provided by the Pittsburgh Supercomputer Center, the Texas Advanced
Computing Center, and the National Institute for Computational Sciences.
MP is supported by Science Foundation Ireland under research grant 07/RFP/PHYF168.
The Chroma software suite\cite{chroma} was used.  We thank our colleagues
within the Hadron Spectrum Collaboration.



\bibliographystyle{aipprocl} 



\end{document}